\documentclass[12pt,english,longbibliography,nofootinbib,superscriptaddress,12pt,sort&compress,showkeys]{article}
\usepackage[latin9]{inputenc}
\usepackage[letterpaper]{geometry}
\usepackage[active]{srcltx}
\usepackage{amsmath}
\usepackage{amssymb}
\usepackage{graphicx}
\usepackage[numbers]{natbib}

\makeatletter

\newcommand{\lyxdot}{.}

\newcommand{\lyxaddress}[1]{
	\par {\raggedright #1
	\vspace{1.4em}
	\noindent\par}
}

\usepackage{aecompl}\usepackage{epstopdf}

\usepackage[raggedrightboxes]{ragged2e}

\usepackage{babel}

\makeatother

\usepackage{babel}
\begin{document}
\title{Atomistic modeling of the structure and diffusion processes at Al(110)/Si(001)
interphase boundaries obtained by vapor deposition}
\author{Yang Li and Yuri Mishin}
\maketitle

\lyxaddress{Department of Physics and Astronomy, MSN 3F3, George Mason University,
Fairfax, Virginia 22030, USA}
\begin{abstract}
\noindent We report on molecular dynamics simulations of the atomic
structure and diffusion processes at Al(110)/Si(001) interphase boundary
created by simulated vapor deposition of Al(Si) alloy onto Si(001)
substrate. An array of parallel misfit dislocations of both full and
partial types is observed at the interface. Si atoms segregate to
the misfit dislocations, with segregation to full dislocations being
stronger. The interface diffusion is dominated by short-circuit diffusion
along the misfit dislocations, creating a significant diffusion anisotropy.
Diffusion of Al and Si atoms along the full misfit dislocations is
faster than along the partial misfit dislocations. Due to the presence
of the misfit dislocations, diffusion at the Al(110)/Si(001) interface
studied here is faster than diffusion at the Al(111)/Si(111) interfaces
investigated in our previous work.
\end{abstract}

\maketitle

\section{Introduction}

Metal-nonmetal interphase boundary (IPB) diffusion controls the rate
of many processes in materials, including phase transformations, diffusional
creep in composite materials, and microstructure development in multi-phase
alloys. Experimental measurements of IPB diffusion are extremely challenging,
scarce, and usually indirect \citep{Straumal:1984aa,Kosinova:2015aa,Kumar:2018aa,Barda:2020aa}.
As a result, the fundamental understanding of IPB diffusion at metal-nonmetal
interfaces remains highly incomplete. Most of the current knowledge
about the IPB diffusion comes from computational modeling and simulations. 

In recent years, Al/Si IPBs have attracted much attention as a model
system to understand the diffusion-controlled interfacial creep in
metal-matrix composites \citep{Funn:1998aa,Dutta:2000aa,Peterson:2003aa,Peterson:2002aa,Peterson:2004aa,wu2022atomistic,Al-Si-IPB,Chesser:2024aa,li2024atomistic}.
Most of the computational studies of Al/Si IPBs were focused on the
interfacial structure, tensile debonding, and interface shearing \citep{wu2022atomistic,Wu:2024aa,Gall:2000aa}.
To our knowledge, mass transport along Al/Si interfaces has only been
addressed in two studies \citep{Al-Si-IPB,li2024atomistic}. Chesser
et al.~\citep{Al-Si-IPB} used a combination of molecular dynamics
(MD) and Monte Carlo simulations to investigate the high-temperature
structures and diffusion of Al and Si at Al/Si interfaces constructed
by either direct bonding or simulated epitaxy. The first method bonds
together Al to Si single crystals with predetermined crystallographic
orientations, while in the second method a liquid Al(Si) alloy is
solidified on a Si substrate. The IPB diffusion was found to be so
slow that only diffusion coefficient at the Al(110)/Si(001) interface
containing disconnections was measurable. Other interfaces were either
unstable or their diffusivity was too slow to be measured using conventional
MD simulations. 

In experiments, Al/Si interfaces are commonly produced by vapor deposition
of an epitaxial Al layer onto a Si substrate with a specific orientation.
Two crystallographic orientations of epitaxial Al/Si structures were
found to be stable at high temperatures: Al(111)/Si(111) \citep{legoues1986atomic,liu2018perfect,mcskimming2017metamorphic,choi1987epitaxial,groger2001ultrathin,tsai2019nano}
and Al(110)/Si(001) \citep{thangaraj1992epitaxial,yamada1984epitaxial}.
Other Al/Si structures have also been reported, but most of them are
unstable at high temperatures or do not produce a high-quality Al
layer. Therefore, to model more realistic interfaces than in \citep{Al-Si-IPB},
we have recently created a set of Al(111)/Si(111) IPBs by simulated
vapor deposition of Al on Si(111) substrates \citep{li2024atomistic}.
The resulting Al(111)/Si(111) interface structures were in good agreement
with experimental observations \citep{legoues1986atomic,liu2018perfect,mcskimming2017metamorphic,choi1987epitaxial,groger2001ultrathin,tsai2019nano}.
That work has demonstrated advantages of the simulated deposition
in capturing the experimentally observed microstructures (such as
threading dislocations and grain boundaries) in epitaxial layers without
relying on \emph{a priori} assumptions about the crystallographic
orientation relationships at the interfaces \citep{li2023dislocation,li2024atomistic}. 

In this study, we investigate Al/Si interfaces obtained by depositing
Al-Si layer onto a Si(001) substrate, closely mimicking the experimental
deposition process. After introducing our simulated methodology in
section \ref{sec:Methodology}, we present a detailed analysis of
the interface structure (section \ref{subsec:Interphase-structure})
and diffusion (section \ref{subsec:Diffusion-coefficients}). In section
\ref{sec:Conclusions}, we formulate our conclusions.

\section{Methodology\label{sec:Methodology}}

\subsection{Vapor deposition simulations}

The MD simulations used the Large-scale Atomic/Molecular Massively
Parallel Simulator \citep{Plimpton95} and the Al-Si interatomic potential
developed by Saidi et al.~\citep{Saidi_2014}. The atomic structures
were visualized and analyzed using the Open Visualization Tool (OVITO)
\citep{Stukowski2010a}. The Si substrate had its $[1\overline{1}0]$,
$[110]$, and $[001]$ crystallographic directions aligned parallel
to the Cartesian axes $X$, $Y$, and $Z$ of the rectangular simulation
box. The substrate dimensions were $21.5\times21.5\times7.5$ nm$^{3}$.
Periodic boundary conditions were applied in the lateral directions
($X$ and $Y$) with a fixed boundary condition along the vertical
$Z$ direction. A 1 nm thick rigid slab at the bottom of the substrate
was fixed. The lattice parameter in the substrate was adjusted to
the thermal expansion coefficient of Si at the growth temperature,
and the atomic velocities were initiated according to this temperature.
A gas of Al and Si atoms with a chosen Al:Si ratio was created well
above the substrate surface. The atomic velocities of the gas atoms
pointed towards the substrate and had magnitudes drawn from the Maxwell
velocity distribution at the growth temperature. 

\begin{figure}
\centering{}\includegraphics[width=0.6\linewidth]{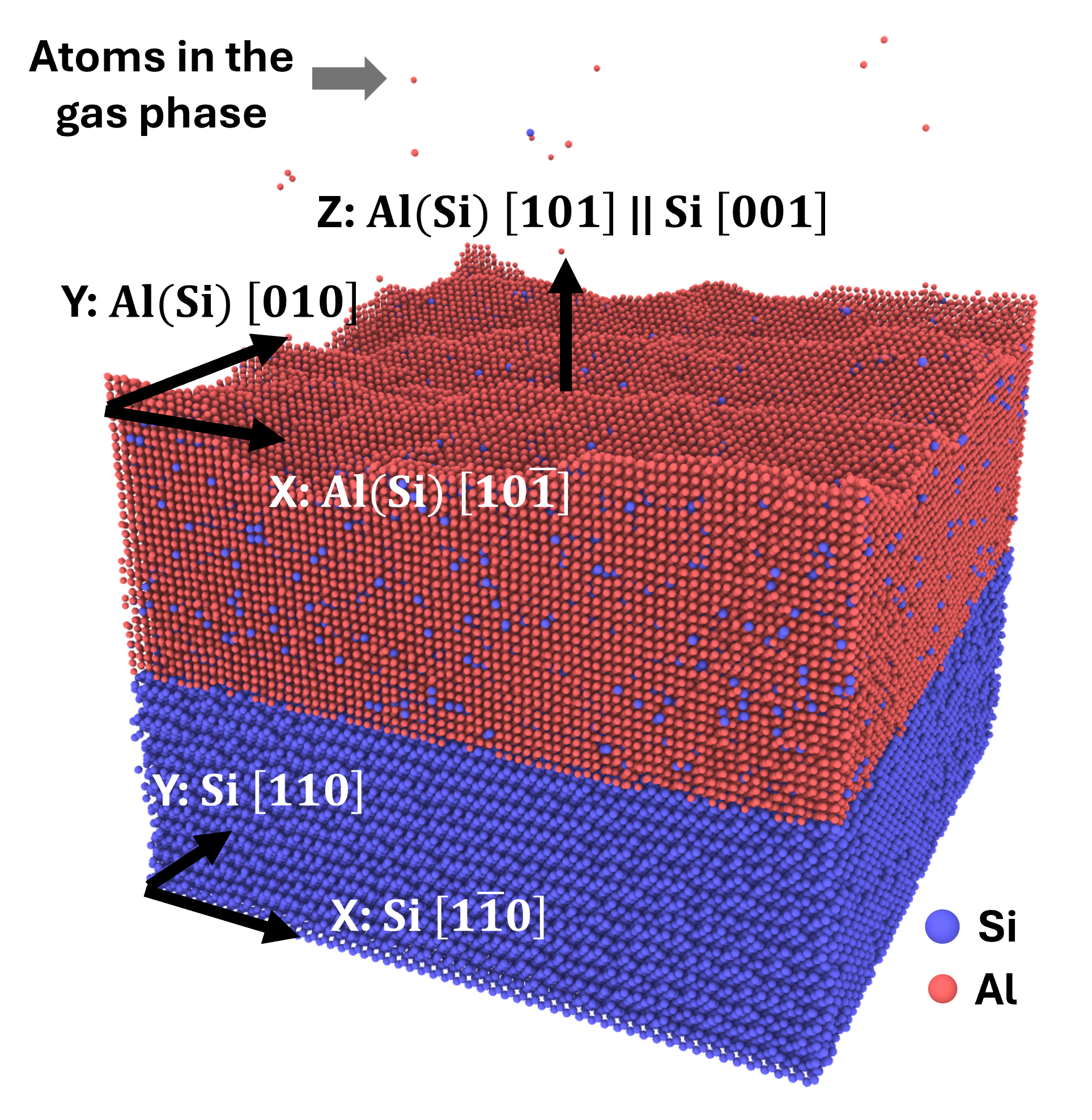}
\caption{Simulated growth of Al$_{0.95}$(Si$_{0.05}$) layer on Si(001) substrate
by vapor deposition at the temperature of 604 K. The Cartesian $X$,
$Y$, and $Z$ directions are along the $[1\overline{1}0]$, $[110]$,
and $[001]$ crystallographic directions of the substrate.}
\label{fig:Model}
\end{figure}

During the MD simulation, the atoms hitting the substrate surface
gradually formed a growing layer of Al-Si alloy, which we denote Al(Si).
The chemical composition of the alloy was controlled by the Al:Si
ratio in the gas phase. At each temperature, the Al(Si) composition
was chosen to be slightly below the Si concentration on the solvus
line of the Al-Si phase diagram calculated in \citep{Al-Si-IPB} using
the same interatomic potential. This ensured that Al(Si) was in a
single-phase state. The deposition was performed at the temperatures
of 578 K, 604 K, 622 K, and 648 K. The latter temperature is close
to the eutectic temperature (approximately 677 K) predicted by the
potential. The respective Si concentrations in Al(Si) were 3.5\%,
5.0\%, 6.8\%, and 9.5\%. (In this work, all chemical compositions
are measured in atomic per cents of Si.) Fig.~\ref{fig:Model} is
an example of the simulation setup for depositing an Al$_{0.95}$(Si$_{0.05}$)
layer on Si(001) substrate at the temperature of 604 K. 

\subsection{Diffusion coefficient calculations}

\begin{figure}
\centering{}\includegraphics[width=0.75\linewidth]{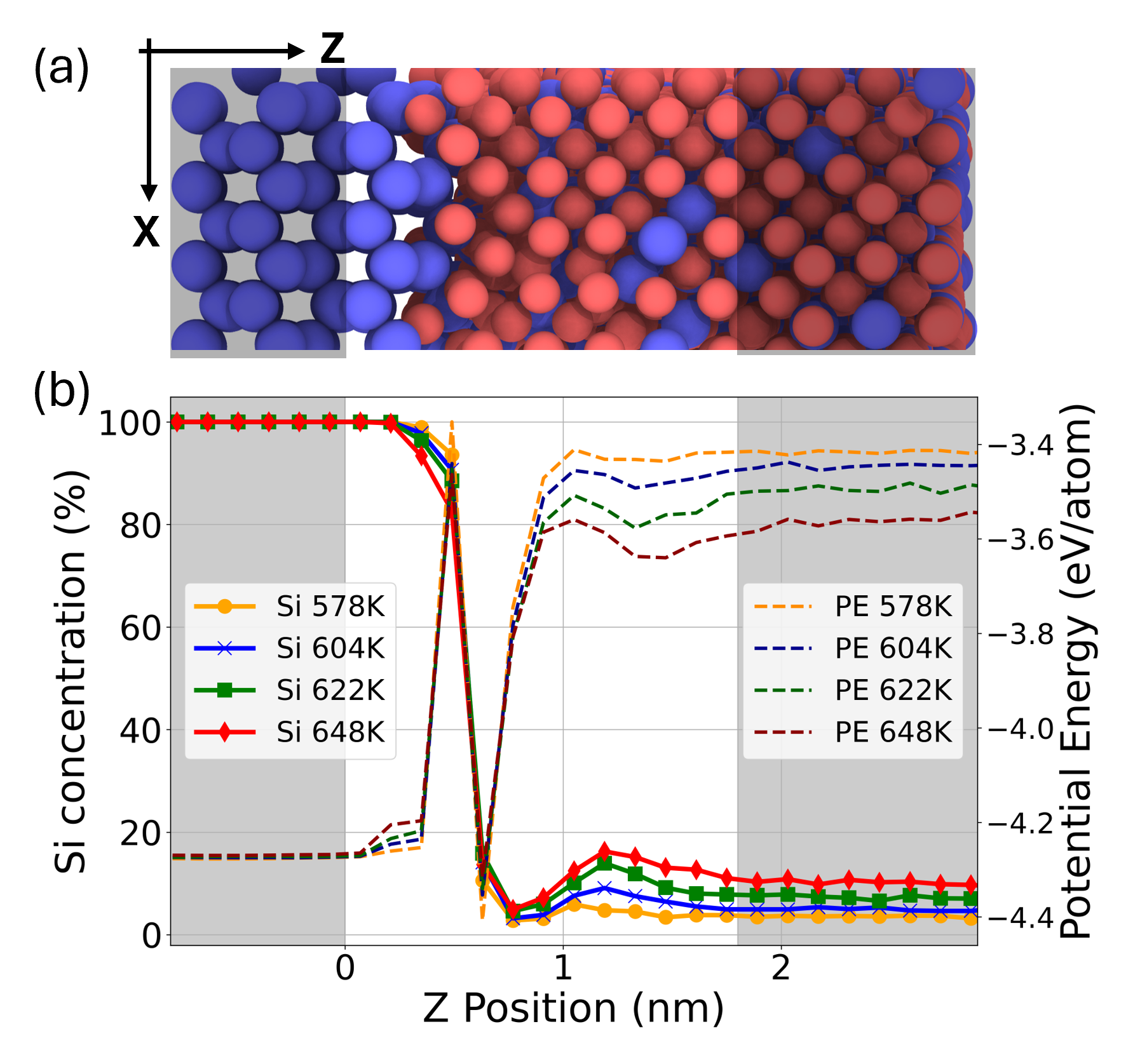}
\caption{(a) The atomic structure of the Al(Si)/Si interface within the $z$
range indicated in (b). (b) Si concentration and atomic potential
energy as a function of $z$ coordinate. The white layer represents
the interface regions, while the gray regions on either side represent
bulk phases. }
\label{fig:Interface region}
\end{figure}

After the deposition, the Al(Si)/Si structure was annealed at the
growth temperature for at least 30 ns by running MD simulations in
the canonical (NVT) ensemble. During the anneal, two 1 nm thick rigid
slabs at the top and bottom of the Al(Si)/Si structure were fixed.
The diffusion coefficients of Al and Si at the IPB were measured by
tracking the motion of atoms within a 1.8 nm thick probe layer centered
at the interface. The thickness of this layer was chosen based on
the potential energy and Si concentration profiles across the Al(Si)/Si
IPB. Specifically, the Al(Si)/Si interface region was divided into
a series of 0.14 nm thick layers parallel to the substrate. The potential
energy of the atoms and the Si concentration in each layer were calculated
as a function of the layer's coordinate $z$. Fig.~\ref{fig:Interface region}
shows that both the potential energy and the Si concentration exhibit
drastic change with position in the interface region and converge
to the respective coordinate-independent values outside the interface
region. The largest variation occurs at $z\approx0.55$ nm. Considering
that the variations are localized approximately between $z=0$ and
$z=1.8$ nm, we took this interval as the probe region for computing
the IPB diffusivities. As in previous work \citep{Chesser:2024aa,Al-Si-IPB,li2024atomistic},
we found that the Si atoms in the substrate were virtually immobile
on the time scale of conventional MD simulations. Therefore, when
calculating the diffusion coefficients of Si atoms, we excluded the
substrate Si atoms defined as those with the $z$ coordinate below
0.55 nm.

As discussed in the next section, misfit dislocations were observed
at the Al(Si)/Si IPB. Diffusion coefficients in the misfit dislocation
cores were computed by tracking the motion of atoms within pipe-shape
probe regions containing the dislocation lines. The pipes were parallel
to the substrate and had their $z$-coordinates at $0.55$ nm (the
center of the interface layer). The pipe radius was chosen to be $1.25$
nm. As above, the Si atoms with $z<0.55$ nm were excluded from the
diffusion calculations.

The diffusion coefficients were calculated from the Einstein relations
$D_{x}=\langle x^{2}\rangle/2t$ and $D_{y}=\langle y^{2}\rangle/2t$,
where $\langle x^{2}\rangle$ and $\langle y^{2}\rangle$ are the
mean squared displacements (MSDs) of atoms in the two in-plane directions,
and $t$ is the simulation time. Only atoms that remained in the probe
region at the initial and final times were included in the MSD calculation.
The calculation required a linear fit to MSD versus time curves, which
we noisy due to thermal fluctuations. Therefore, prior to fitting,
the curves were subjected to bootstrap resampling to smooth the noise
and estimate the errors. The hyper-parameters chosen for the resampling
included a block length of 0.1 ns and a number of resampled trajectories
equal to 100.

\section{Results}

\subsection{Interphase boundary structure\label{subsec:Interphase-structure}}

The simulations have shown that the deposited Al(Si) layer exhibits
the same crystallographic orientation relationship with the Si(001)
substrate regardless of the growth temperature (respectively, the
Si concentration). The orientation relationship is: $X$: Al(Si)$[10\overline{1}]\parallel$
Si$[1\overline{1}0]$, $Y$: Al(Si)$[010]\parallel$ Si$[110]$, and
$Z$: Al(Si)$[101]\parallel$ Si$[001]$. This orientation relationship
is consistent with experimental observations \citep{thangaraj1992epitaxial,yamada1984epitaxial}.
We refer it as simply ${\text{Al(Si)}\{110\}}\parallel{\text{Si}}\{001\}$. 

\begin{figure}
\centering{}\centering \includegraphics[width=1\linewidth]{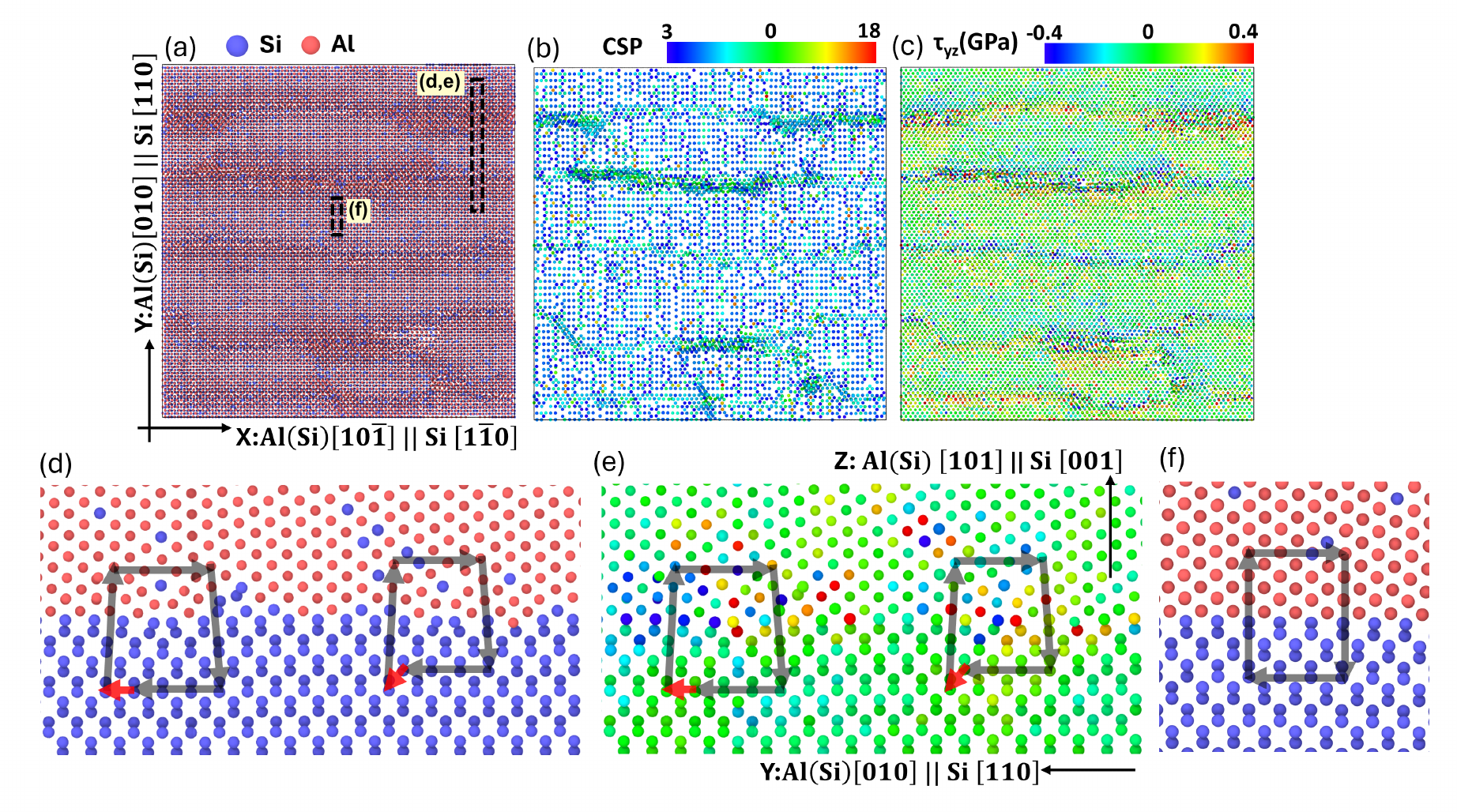}
\caption{(a) Plane view of the atomic structure of the Al$_{0.95}$(Si$_{0.05}$)/Si
IPB grown at the temperature of 604 K. (b) Interfacial Al atoms color-coded
by CSP. (c) Interfacial Al atoms color-coded by the $\tau_{yz}$ stress
component. (d) Side view of a full (left) and a partial (right) misfit
dislocation in the region marked by the dashed rectangle in (a). The
Burgers circuits are drawn around the two dislocations to demonstrate
the closure failure. (e) Same as (d) but the atoms are color-coded
by the $\tau_{yz}$ stress component. (f) Coherent interface region
marked by the dashed rectangle in (a), showing Burgers circuit closure.}
\label{fig:Misfit dislocations}
\end{figure}

\begin{figure}
\centering{}\centering \includegraphics[width=0.75\linewidth]{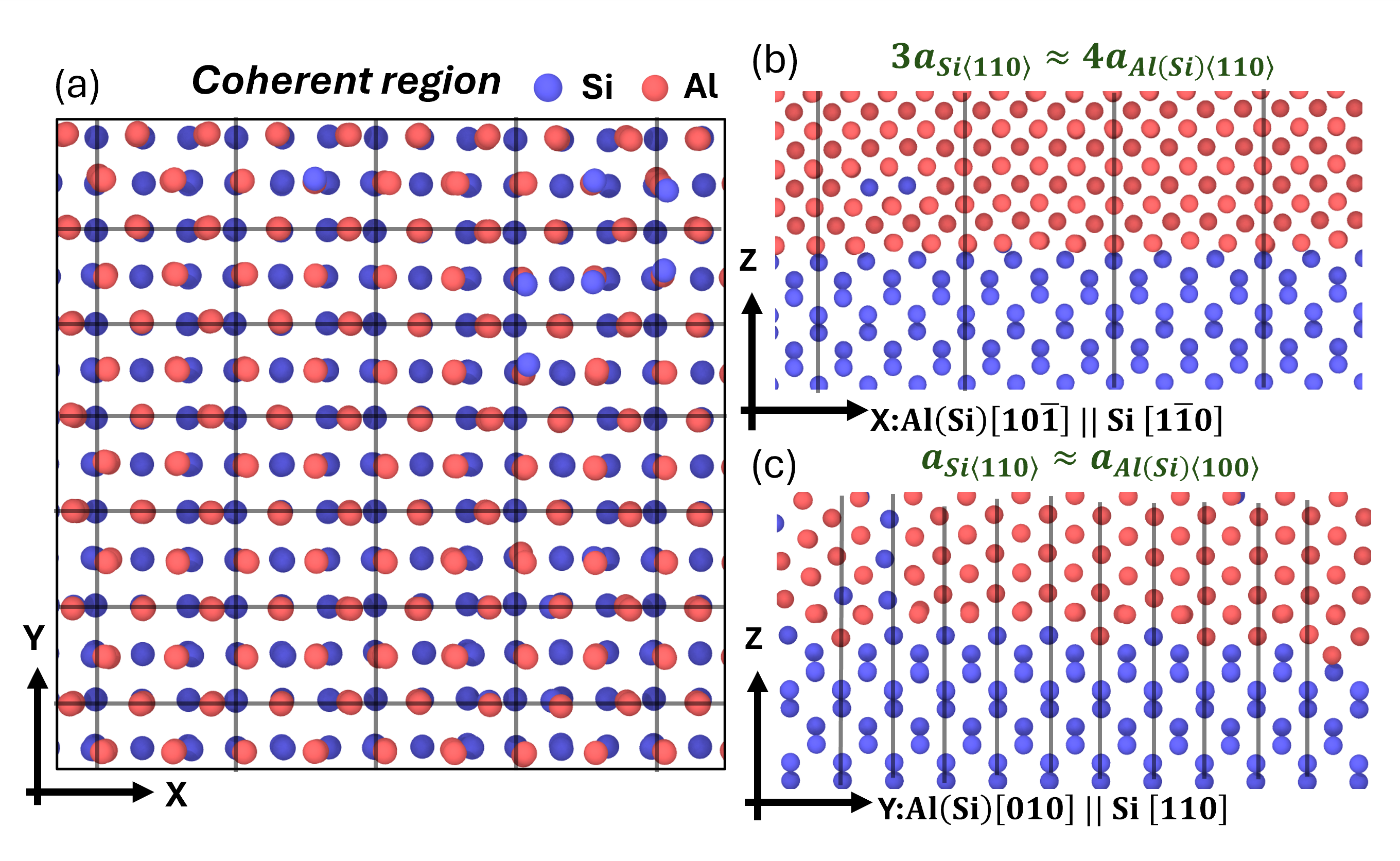}
\caption{(a) Plane view and (b,c) side views of a coherent region at the Al$_{0.95}$(Si$_{0.05}$)/Si
interface. The black lines indicate the periodicity of the coherent
region.}
\label{fig:Coherent region}
\end{figure}

The IPB contains an array of misfit dislocations separated by interface
regions of high coherency (Fig.~\ref{fig:Misfit dislocations}(a)).
To reveal the misfit dislocations in the MD snapshots, we applied
two criteria: the local excess of stress and the centrosymmetry parameter
(CSP) \citep{Kelchner98}. The latter is commonly used to characterize
lattice distortions. Both quantities were computed and visualized
with OVITO \citep{Stukowski2010a}. An example of visualization of
misfit dislocations by these criteria is shown in Fig.~\ref{fig:Misfit dislocations}(b,c).
The dislocations are aligned parallel to the $X$ direction, which
can be explained by the anisotropy of the lattice mismatch. Indeed,
taking the Al$_{0.95}$(Si$_{0.05}$)/Si structure as an example,
the $[1\overline{1}0]$ direction of the Si substrate and the $[10\overline{1}]$
direction of the Al(Si) layer are parallel to the $X$ direction and
their lattice spacings are 0.385 nm and 0.291 nm, respectively. The
$[110]$ direction of the Si substrate and the $[010]$ direction
of the Al(Si) layer are parallel to the $Y$ direction and their lattice
spacings are 0.385 nm and 0.412 nm, respectively. In the coherent
IPB regions between the misfit dislocations, the periodicity along
the $X$ direction is every 4 $(10\overline{1})$ planes of Al(Si)
and every 3 $(1\overline{1}0)$ planes of Si (Fig.~\ref{fig:Coherent region}a
and \ref{fig:Coherent region}b). In the $Y$ direction, the periodicity
is every $(010)$ plane of Al(Si) and every $(110)$ plane of Si (Fig.~\ref{fig:Coherent region}a
and \ref{fig:Coherent region}c). With these periodicities, the lattice
mismatches between the Al(Si) layer and the substrate are about 0.8\%
along the $X$ direction and about 6.8\% along the $Y$ direction.
Here, the mismatch $\delta$ is defined as the normalized difference
between the lattice periodicities $l_{Si}$ and $l_{Al(Si)}$ in the
respective directions: $\delta=2(l_{Si}-l_{Al(Si)})/(l_{Si}+l_{Al(Si)})$
\citep{zhu2005misfit}. Since the mismatch along the $X$ direction
is relatively small and along the $Y$ direction much larger, this
explains why the misfit dislocation lines are nearly parallel to the
$X$ direction. 

Although these estimates utilized the lattice parameters of the Al$_{0.95}$(Si$_{0.05}$)/Si
system at 604 K, the conclusion about the large mismatch difference
between the two directions remains valid for the lattice parameters
at other temperatures of the present simulations. However, the temperature
does affect the dislocation density. As demonstrated in Supplementary
Fig.~\ref{fig:Interface structures}, the spacing between the dislocation
lines increases with temperature. 

Because the crystal structures of the Al(Si) phase and the Si substrate
are different, the Burgers vectors of the misfit dislocation could
not be determined automatically using the DXA algorithm of OVITO \citep{Stukowski2010a}.
Instead, the Burgers vector were obtained manually by drawing Burgers
circuits around the misfitting (incoherent) regions, as demonstrated
in Fig.~\ref{fig:Misfit dislocations}(d,e). This procedure is similar
to the one for lattice dislocations \citep{anderson2017theory,hull2011introduction}
and applied to phase interfaces \citep{gutekunst1997atomic}. The
coherent interface region required for the construction is shown in
Fig.~\ref{fig:Misfit dislocations}(f). Two types of misfit dislocations
were found: full misfit dislocations with Burgers vectors parallel
to the $Y$ direction, $\mathbf{b}_{\mathrm{full}}=a_{\mathrm{Si}}\left\langle 110\right\rangle /2$,
and partial misfit dislocations with Burgers vectors composed of both
$Y$ and $Z$ components, $\mathbf{b}_{\mathrm{partial}}=a_{\mathrm{Si}}\left\langle 110\right\rangle /4+a_{\mathrm{Si}}\left\langle 100\right\rangle /2$.
Here, $a_{\mathrm{Si}}$ is the cubic lattice constant of Si. The
Si lattice was used as the reference frame because its lattice undergoes
smaller distortions than the lattice of Al(Si).

The fraction of full dislocations increases with temperature. For
example, most dislocations in the Al$_{0.965}$(Si$_{0.035}$)/Si
interface grown at 578 K are partial type, while most dislocations
in the Al$_{0.905}$(Si$_{0.095}$)/Si interface grown at 648 K are
full type. This is illustrated in Supplementary Fig.~\ref{fig:578K 648K misfit dislocation}.
As full dislocations are more effective in relieving the mismatch
strain, fewer of them are required for the same amount of misfit.
This explains why the interfaces contain fewer misfit dislocations
at higher temperatures.

In the experimental study by Thangaraj et al.~\citep{thangaraj1992epitaxial},
nearly-periodic contrast was observed at the Al(110)/Si(001) interface.
The authors suggested that this contrast could be caused by amorphous
pockets or islands. Our results offer an alternative explanation:
the observed periodic patterns could have been signatures of misfit
dislocations or similar metastable structures relieving the mismatch
strain in a manner similar to misfit dislocations.

\begin{figure}
\centering{}\centering \includegraphics[width=0.5\linewidth]{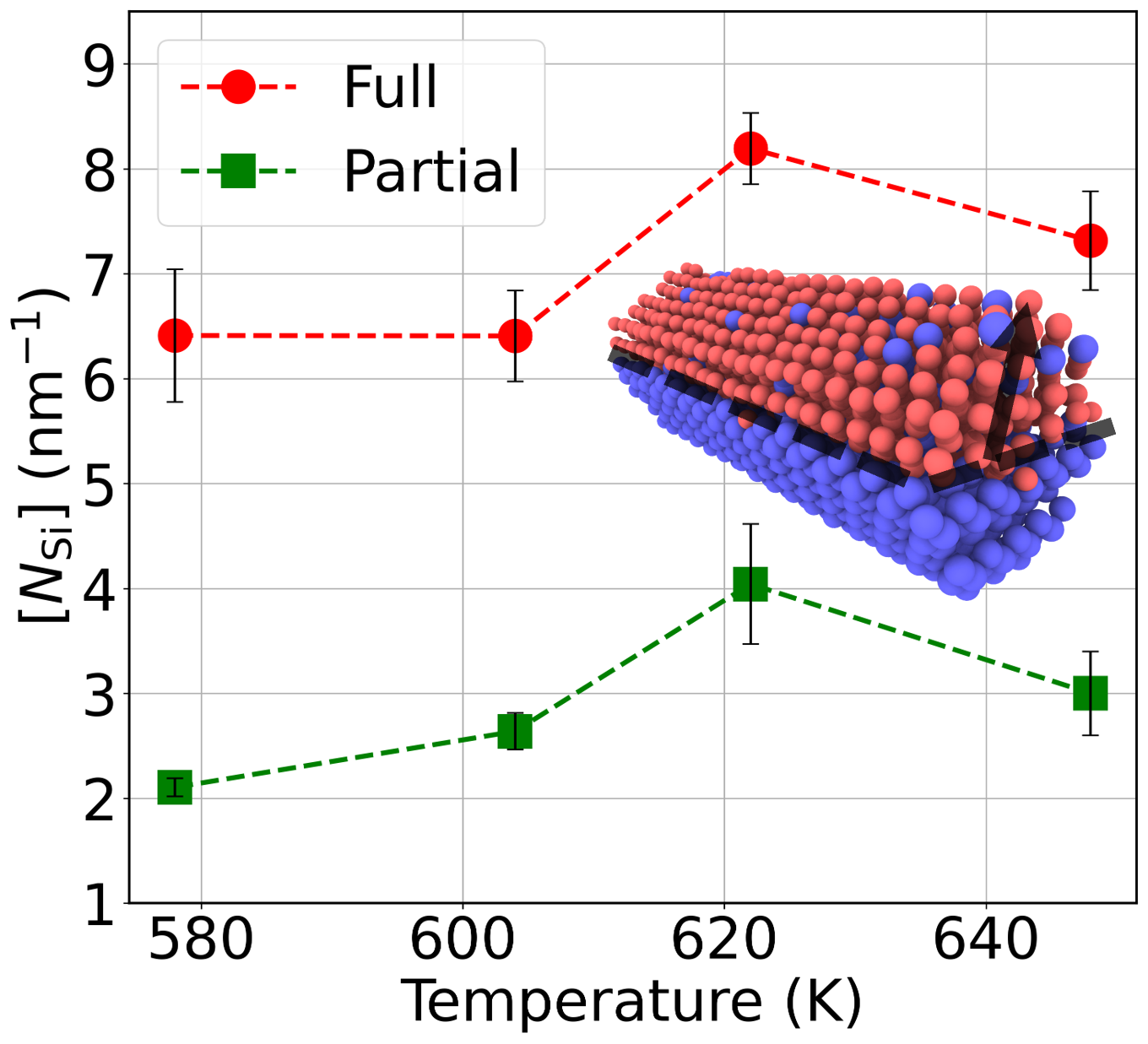}
\caption{Si segregation at the full and partial misfit dislocations in Al(Si)/Si
interface grown at different temperatures. The inset shows a representative
pipe region containing a misfit dislocation, which is used for the
Si segregation calculations. The error bars represent one standard
deviation from averaging over multiple snapshots generated during
MD simulations.}
\label{fig:Segregation}
\end{figure}

In previous computational studies \citep{Chesser:2024aa,Al-Si-IPB,Legros1646},
Si was found to segregate to lattice dislocations in Al . Here, we
found that Si also segregates to misfit dislocations at the Al(110)/Si(001)
IPBs. The amount of segregation was quantified by the excess number
of Si atoms per unit length of the misfit dislocation pipes similar
to those used in diffusion calculations. Specifically, the amount
of segregation $[N_{Si}]$ was defined by $[N_{Si}]=(N_{Si}-Nc)/L$,
where $N$ and $N_{Si}$ denote the total number of atoms and the
number of Si atoms within a pipe region (excluding those in the substrate),
$c$ is the fraction of Si atoms in the lattice inside the Al(Si)
layer, and $L$ is the dislocation pipe length \citep{Koju:2020ab}.
In Fig.~\ref{fig:Segregation}, we plot the amount of Si segregation
at misfit dislocations at the four temperatures studied here. The
plot reveals stronger segregation at full misfit dislocations than
at partial misfit dislocations. No clear correlation with temperature
is observed in the given temperature interval. We note that at a fixed
lattice concentration of the solute, segregation is usually weakens
with increasing temperature. However, in our case, the solute concentration
in the Al(Si) lattice increases with temperature, creating a competing
effect that conceals the expected decrease with temperature.

Although the interfaces are atomically sharp, a small number of Al
atoms intermixes with the top atomic layer ($z<0.55$ nm) of the Si
substrate during the early stages of the vapor deposition process.
Specifically, at the temperatures of 578 K, 604 K, 622 K, and 648
K, the surface density of such atoms is 0.47, 0.71, 0.95, and 1.53
$\text{nm}^{-2}$, respectively. The degree of intermixing increases
with temperature, reflecting the greater mobility of Al and Si atoms
at higher temperatures. Supplementary Fig.~\ref{fig:Intermixing}
demonstrates that the intermixing is localized near the misfit dislocations.
As shown in Fig.~\ref{fig:Interface region}, the opposite intermixing,
with Si atoms diffused into the Al(Si) phase, is also small but extends
over a wider zone above the substrate. 

\subsection{Diffusion coefficients\label{subsec:Diffusion-coefficients}}

\begin{figure}
\centering \includegraphics[width=1\linewidth]{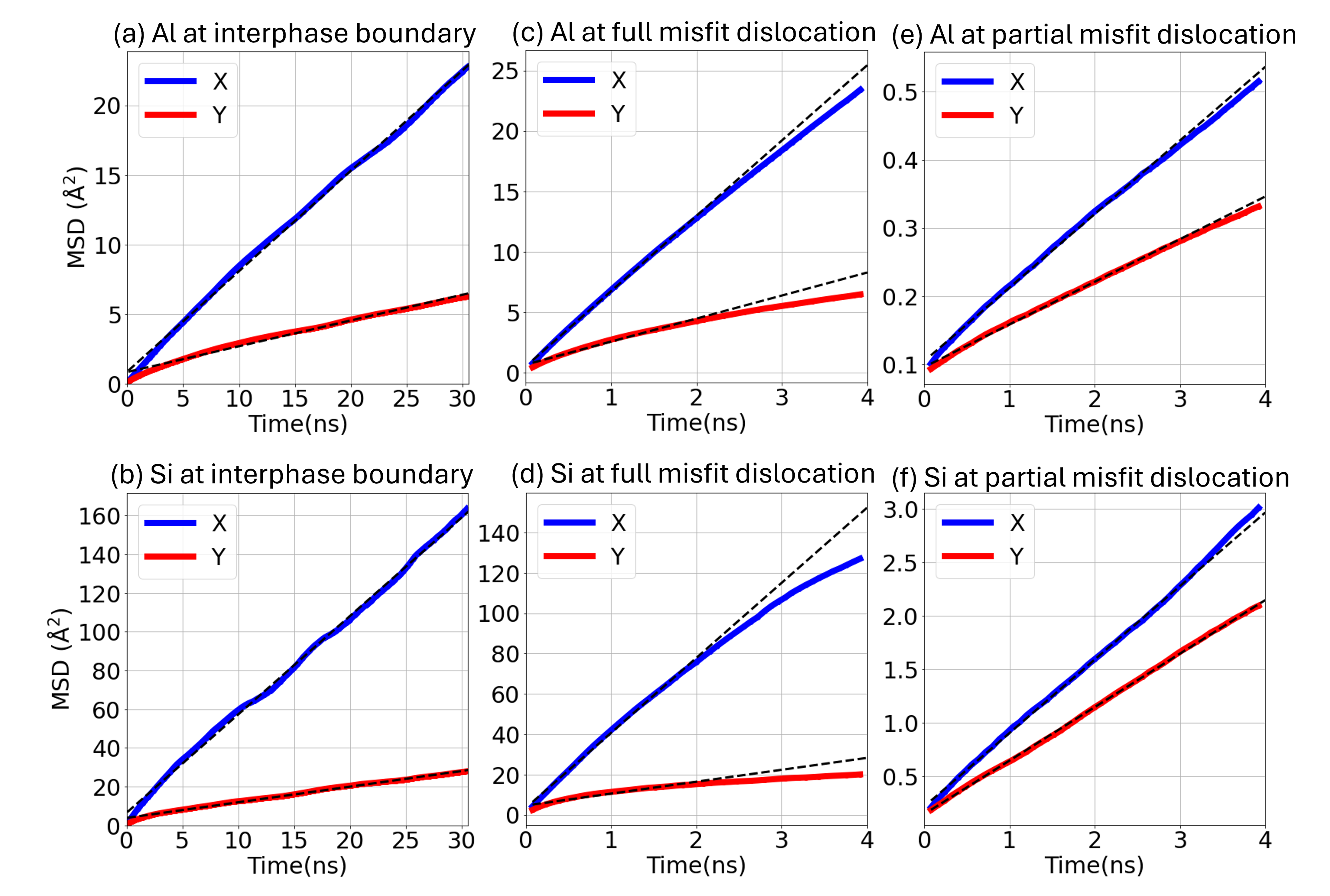}
\caption{Representative MSD versus time plots obtained by MD simulations. Al
and Si atoms diffusing in the $X$ and $Y$ directions at (a,b) the
interphase boundary, (c,d) a selected full misfit dislocation segment,
and (e,f) a selected partial misfit dislocation segment in the Al$_{0.905}$(Si$_{0.095}$)/Si
system. (Si atoms in the substrate were excluded). The temperature
is 648 K.}
\label{fig:MSD}
\end{figure}

Typical MSD versus time plots are shown in Fig.~\ref{fig:MSD}. In
most cases, the linear relationship predicted by the Einstein formula
is followed fairly well. However, deviations from linearity are observed
in some cases. The possible sources of such deviations were discussed
elsewhere \citep{li2024atomistic}. When a plot deviated from linear
behavior, we used its initial (short-time) portion to extract the
diffusion coefficient.

\begin{figure}
\centering \includegraphics[width=1\linewidth]{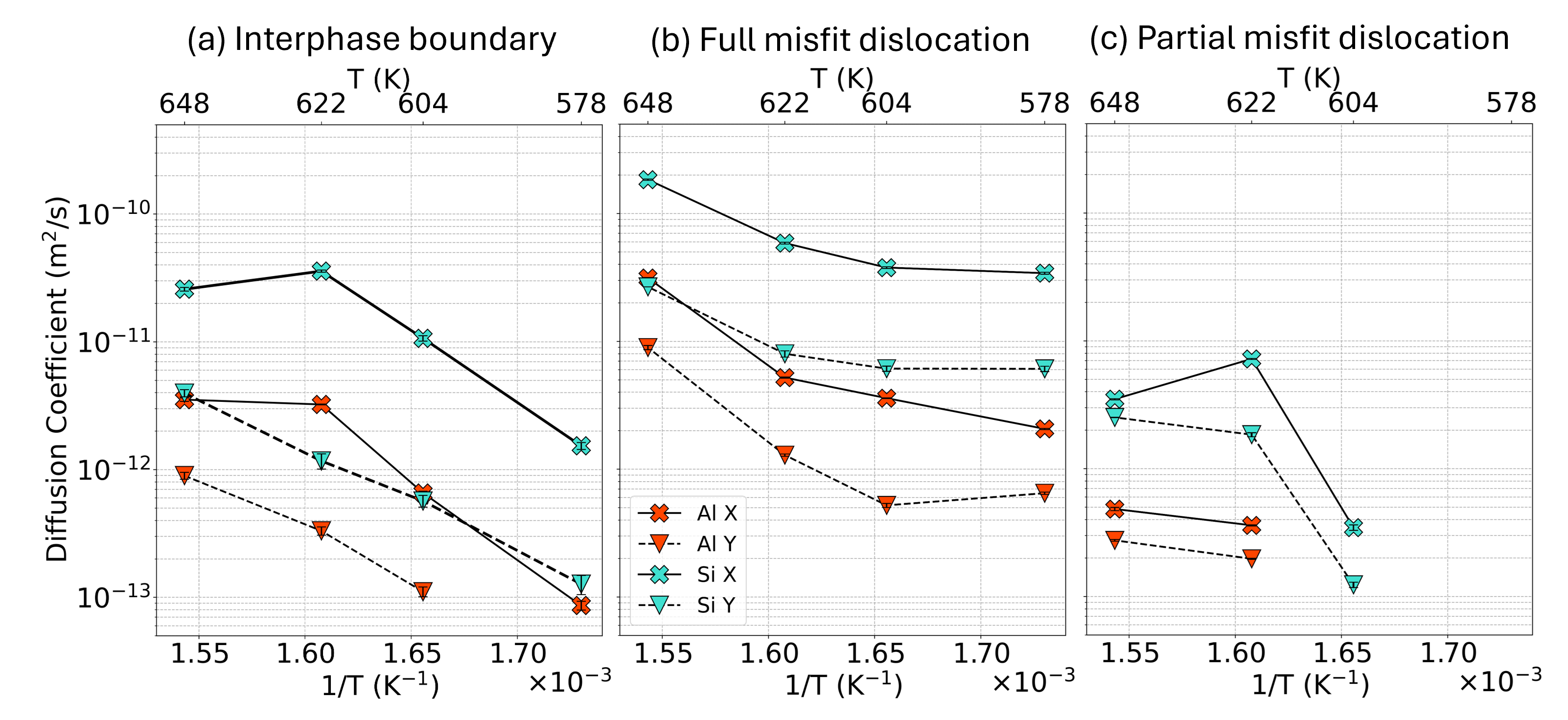}
\caption{Arrhenius diagrams of Al and Si diffusion in the $X$ and $Y$ directions
at (a) Al(Si)/Si interphase boundaries, (b) full misfit dislocations,
and (c) partial misfit dislocations. }
\label{fig:Arrhenius diagram}
\end{figure}

The diffusion coefficients of Al and Si at the IPBs and misfit dislocations
are shown on the Arrhenius diagram (log diffusivity versus inverse
temperature) in Fig.~\ref{fig:Arrhenius diagram}. Recall that the
Si diffusion coefficients represent diffusion on the Al(Si) side of
the interface, with the immobile substrate atoms excluded from the
MSD calculation. The diffusion coefficients tend to increase with
temperature, but this dependence is noisy and nonlinear. The significant
statistical scatter originates from the limited statistics of the
atomic displacements. Indeed, only a small fraction of the atoms present
in the system are located in the probe regions where their trajectories
are tracked during the MD simulations. Linearity cannot be expected
either. Arrhenius plots are only linear when the process is kinetically
controlled by a simple single-barrier mechanism. In the present case,
the interface region is structurally heterogeneous with multiple diffusion
pathways controlled by different mechanisms \citep{li2024atomistic}.
In addition, the changes in temperature are accompanied by corresponding
changes in the chemical composition and the misfit dislocation density.

The plot in Fig.~\ref{fig:Arrhenius diagram} shows Si atoms exhibit
higher diffusivity than Al atoms both in the IPBs and along the misfit
dislocations. This is consistent with our previous study of the Al(Si)(111)/Si(111)
systems \citep{li2024atomistic}, in which Si atoms exhibit a higher
mobility than Al atoms not only at IPBs but also in grain boundaries
and lattice dislocations. 

Diffusion along the dislocation lines (approximately parallel to the
$X$ direction) is faster than diffusion in the direction normal to
the dislocation lines (approximately along the $Y$ direction) (Fig.~\ref{fig:Arrhenius diagram}(b,c)).
This strong diffusion anisotropy is a clear manifestation of the short-circuit
diffusion effect in dislocation cores \citep{Chesser:2024aa}. Note
that diffusion along full misfit dislocations is faster than diffusion
along partial misfit dislocations. The difference in diffusion coefficients
can be up to two orders of magnitude. 

The collective diffusion data obtained in this study shows that Si
diffusion along the full misfit dislocations is the fastest process
in the Al(110)/Si(001) interfaces, while Al diffusion normal to the
dislocation lines is the slowest process. Diffusion in the coherent
interface regions is even slower. The respective diffusion coefficients
do not appear on the Arrhenius diagram because they are below the
resolution of our method. For this reason, the IPB diffusivities shown
in Fig.~\ref{fig:Arrhenius diagram}(a) are strongly dominated by
diffusion along the misfit dislocations. 

\section{Conclusions\label{sec:Conclusions}}

In this study, we have employed physical vapor deposition simulations
to construct Al(Si)/Si(100) IPBs at high temperatures ranging from
578 K to 648 K. The deposited Al(Si) alloy phase crystallizes with
the orientation relationship Al(Si)$[10\overline{1}]\parallel$ Si$[1\overline{1}0]$,
Al(Si)$[010]\parallel$ Si$[110]$, and Al(Si)$[101]\parallel$ Si$[001]$
in agreement with experimental observations \citep{thangaraj1992epitaxial,yamada1984epitaxial}.
The Al(110)/Si(001) IPB obtained contains an array of full and partial
misfit dislocations parallel to the Al(Si)$[10\overline{1}]\parallel$
Si$[1\overline{1}0]$ direction. This alignment of the dislocation
lines is explained by the strong anisotropy of the lattice mismatch
between the Al(Si) and Si phases. The fraction of partial-to-full
dislocations and the total misfit dislocation density decrease with
increasing temperature. Si atoms segregate to the misfit dislocations,
with the amount of segregation being larger for full dislocations.

The IPB diffusion is dominated by short-circuit diffusion along the
misfit dislocations, creating a strong diffusion anisotropy. Diffusion
of both Al and Si atoms along full misfit dislocations is faster than
diffusion along partial misfit dislocations. Si diffusion along full
misfit dislocations is the fastest diffusion mechanism observed in
this study. Overall, the diffusion mobility in the Al(110)/Si(001)
IPB studied here is higher than in the Al(111)/Si(111) IPBs studied
in our previous work \citep{li2024atomistic}.

\bigskip{}

\noindent \textbf{Acknowledgments}

This research was supported by the U.S. Department of Energy, Office
of Basic Energy Sciences, Division of Materials Sciences and Engineering,
under Award \# DE-SC0023102.


\newpage{}

\newpage{}

\clearpage{}

\appendix
\begin{center}
\textbf{SUPPLEMENTARY INFORMATION}
\par\end{center}

\bigskip{}
\bigskip{}

\global\long\def\thefigure{S\arabic{figure}}%
 \setcounter{figure}{0}

\begin{figure}[h]
\centering \includegraphics[width=1\linewidth]{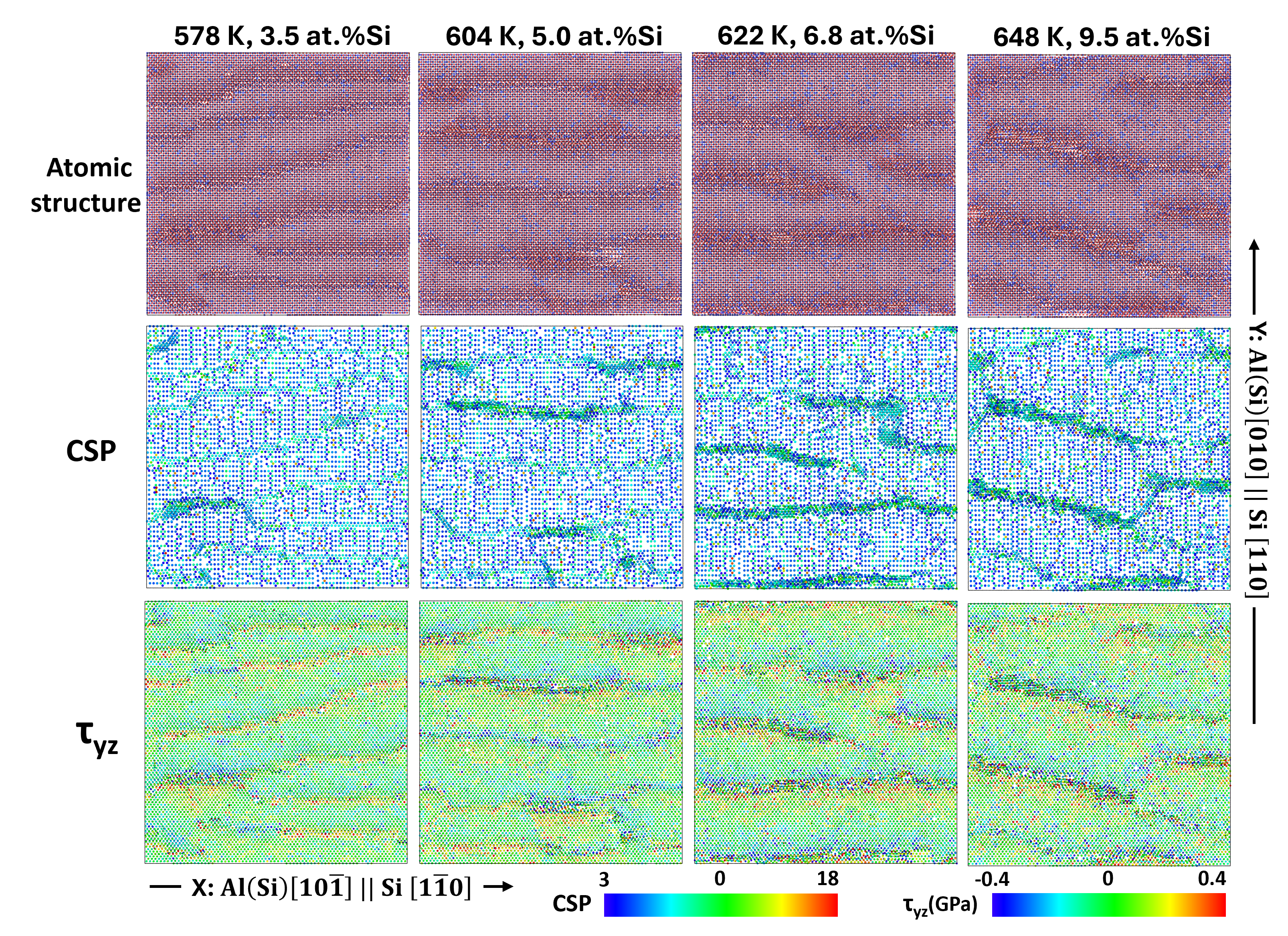}
\caption{Plane views of the atomic structures (upper row), centrosymmetry parameter
(CSP) (middle row), and shear stress $\tau_{yz}$ (bottom row) of
Al(Si)/Si IPBs in systems grown at the temperatures of 578 K (first
column), 604 K (second column), 622 K (third column), and 648 K (fourth
column). The Si concentrations in the Al(Si) phase are 3.5\%, 5.0\%,
6.8\%, and 9.5\%, respectively. }
\label{fig:Interface structures} 
\end{figure}

\begin{figure}[h]
\centering \includegraphics[width=1\linewidth]{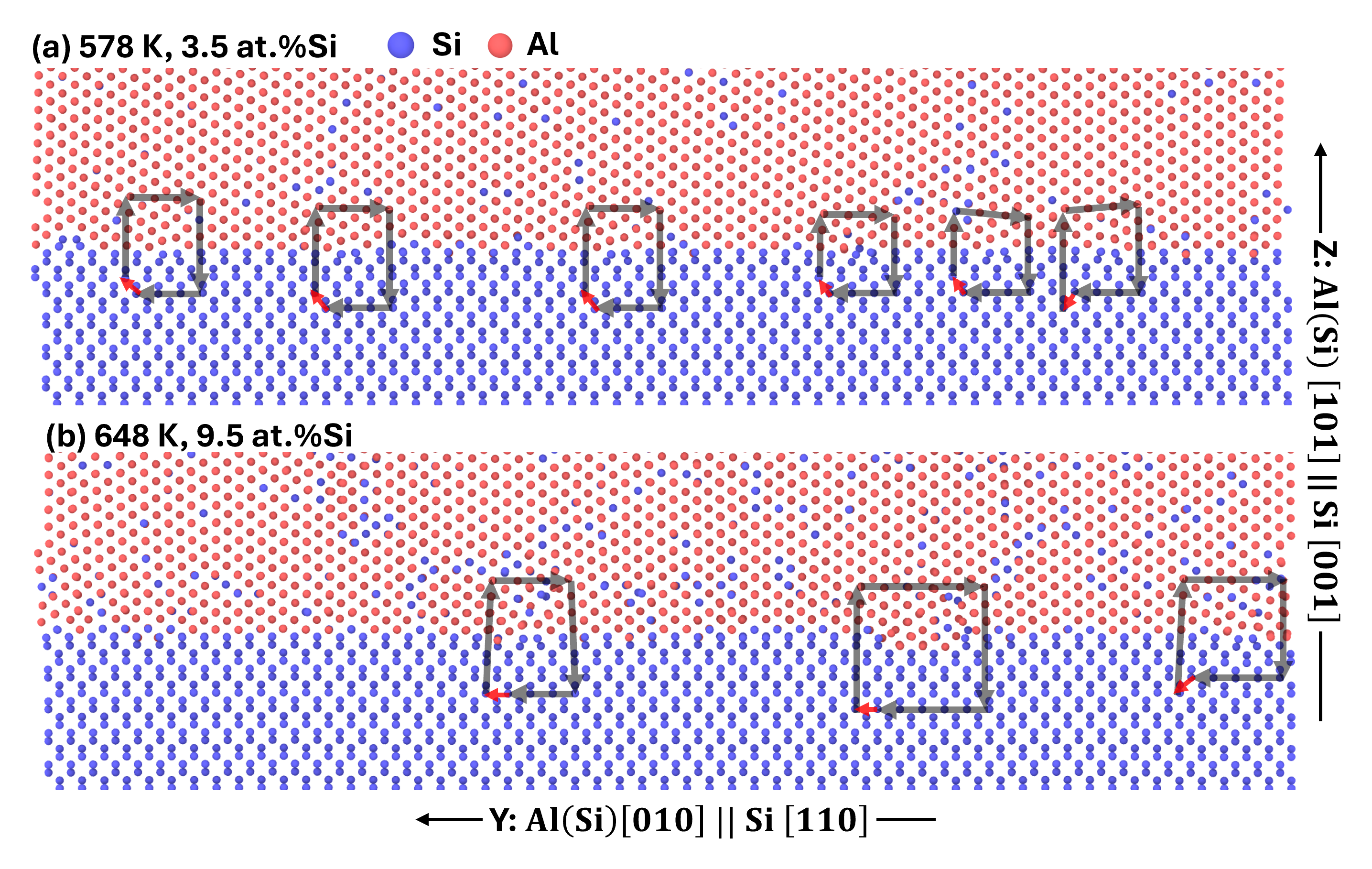}
\caption{Side views of selected interface regions in the Al(Si)/Si structures
grown at the temperatures of (a) 578 K and (b) 648 K. Burgers circuits
are drawn around the misfit dislocations at the interface to characterize
the dislocation types. All dislocations in (a) and the rightmost dislocation
in (b) are partial type, while all other dislocations are full type.}
\label{fig:578K 648K misfit dislocation} 
\end{figure}

\begin{figure}[h]
\centering{}\centering \includegraphics[width=1\linewidth]{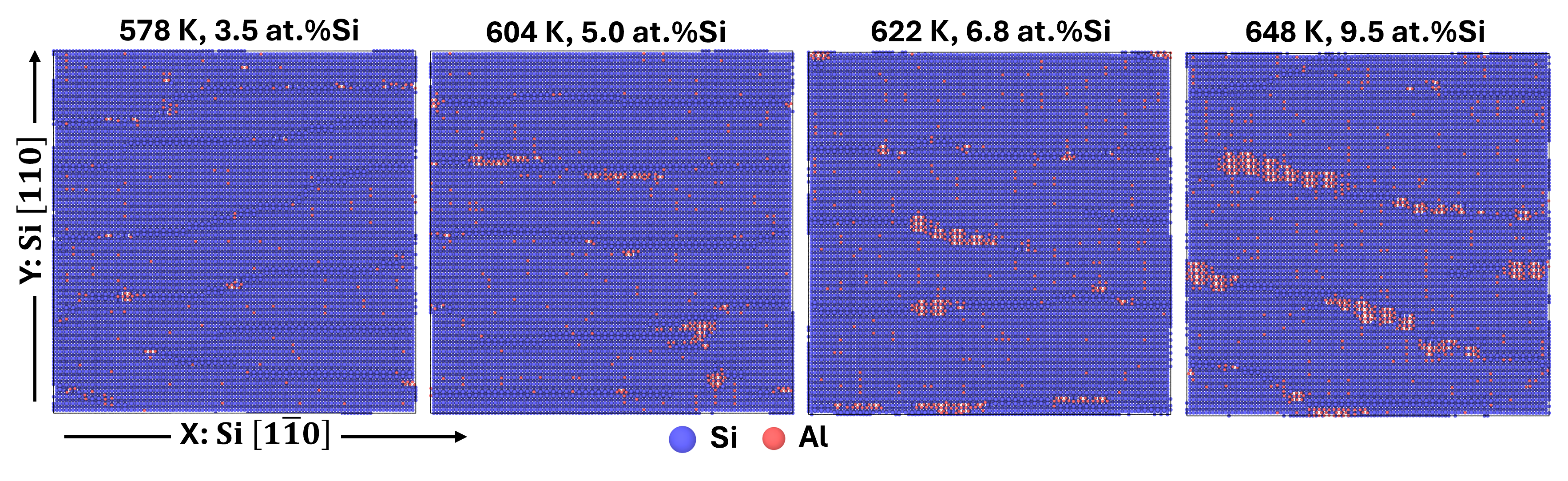}
\caption{Plane views of the top layer of the Si(001) substrates in the Al(Si)/Si
structure grown at temperatures from 578 K to 648 K. The respective
fractions of intermixed Al atoms per interface area are 0.47, 0.71,
0.95, 1.53 $\text{nm}^{-2}$.}
\label{fig:Intermixing} 
\end{figure}

\end{document}